\documentclass[preprint]{elsarticle}
\usepackage[utf8]{inputenc}

\usepackage{natbib}
\usepackage{amsmath}
\usepackage{amsfonts}
\usepackage{graphicx}
\usepackage{microtype}
\usepackage{hyperref}
\usepackage{xcolor}
\usepackage{algorithm}
\usepackage{algpseudocode}

\newcommand{\myref}[1]{\textbf{\nameref{#1}}}
\newcommand{\edit}[1]{{#1}}

\begin{document}

\title{Simulating Structural Plasticity of the Brain more Scalable than Expected\tnoteref{tref}}

\author[1]{Fabian Czappa\corref{cor1}}
\ead{fabian.czappa@tu-darmstadt.de}

\author[2]{Alexander Geiß}
\ead{alexander.geiss1@tu-darmstadt.de}

\author[3]{Felix Wolf}
\ead{felix.wolf@tu-darmstadt.de}

\tnotetext[tref]{ DOI of published journal article: \href{https://doi.org/10.1016/j.jpdc.2022.09.001}{10.1016/j.jpdc.2022.09.001}.\\
	\copyright{} 2022. This manuscript version is made available under the CC-BY-NC-ND 4.0 license \url{https://creativecommons.org/licenses/by-nc-nd/4.0/}.
}
\cortext[cor1]{Corresponding author}
\fntext[fn1]{Laboratory for Parallel Programming, Department of Computer Science, Technical University of Darmstadt, Germany}

\date{June 2022}

\begin{abstract}
Structural plasticity of the brain describes the creation of new and the deletion of old synapses over time. 
Rinke et al. (JPDC 2018) introduced a scalable algorithm that simulates structural plasticity for up to one billion neurons on current hardware using a variant of the Barnes–Hut algorithm. 
They demonstrate good scalability and prove a runtime complexity of $O(n \log^2 n)$. 
In this comment paper, we show that with careful consideration of the algorithm \edit{and a rigorous proof}, the theoretical runtime can even be classified as $O(n \log n)$.
\end{abstract}

\maketitle


\section*{Introduction}
In a recent publication~\cite{Rinke2018}, the authors presented an algorithm that allows fast and scalable simulations of structural plasticity in the human brain.
They achieved this by formulating the Model of Structural Plasticity (MSP)~\cite{Butz2013} as an N-body problem and adapting the Barnes--Hut-algorithm~\cite{Barnes1986}.

In this comment paper, we want to improve the calculations of the runtime complexity from $O(n \cdot \log^2(n))$ to $O(n \cdot \log(n))$ \edit{by revisiting the theoretical considerations for the same algorithm as in~\cite{Rinke2018}}.
We summarize the algorithm in \myref{sec:adapted}
before calculating the sharp runtime bound in \myref{sec:runtime}.
\edit{Then we derive the \myref{sec:parallel-case} and compare it to the empirical performance models of~\cite{Rinke2018}.}

\section*{Adapted Barnes--Hut-Algorithm}
\label{sec:adapted}
The Barnes--Hut-algorithm allows approximating the force that $N$ bodies exert on another body in $O(\log(n))$ time.
It achieves this by building an octree (in 3D; a quadtree in 2D) of all bodies, dividing the simulation space into halves along each axis until at most one body is left per sub-space.
Then, the algorithm allows approximating the force of a whole subspace by the root of the corresponding subtree whenever the subspace is far away for its size.

The position of a virtual body (an inner node of the tree) is the weighted average of the positions of the approximated real bodies (we call that the \textit{centroid}), weighted by a body-dependent scalar (in our case, the number of vacant synaptic elements).
Whenever this virtual body is used, its weight is the sum of the weights of all approximated bodies.
The approximation is valid if the maximum length of a subspace $l$ divided by the distance from the 
target to the virtual body $d$ is smaller than a previously fixed $\theta > 0$.

If a neuron wants to create a new synapse, it searches for another neuron with a vacant synaptic element of the opposite type (a synapse is always formed from an axon to a dendrite).
It does so by checking the Acceptance Criterion (AC) $l/d < \theta$ for the root of the octree, unpacking a virtual neuron if it does not satisfy the AC, and rechecking its children.
Once a neuron establishes a list of potential partners (actual neurons or virtual neurons that are far enough away), it uses this list to calculate the attracting forces and picks a partner accordingly.
If this partner is a virtual neuron, the searching neuron needs to unpack it and use the Barnes--Hut-algorithm again, this time starting from the virtual neuron instead of the root of the octree.

If the octree is balanced, its height is $O(\log(n))$ for $n$ neurons, and in the worst case, there must be $O(\log(n))$ applications of the Barnes--Hut-algorithm per neuron searching a partner.
This is the case if a neuron repeatedly picks a virtual neuron as close as possible to the respective root.
Overall, the runtime is in $O(n \cdot \log(n) \cdot \log(n))$, i.e., every neuron ($n$) has to perform $O(\log(n))$ Barnes--Hut-algorithms, which all have complexity $O(\log(n))$.

\edit{In the parallel version of this algorithm, each MPI rank is responsible for a fixed number of neurons.
	A rank gets assigned a sub-space in which it places all its neurons and for which it builds such an octree (enforcing a suitable number of MPI ranks to avoid boundary issues).
	In every update step, a rank updates the centroids and the numbers of vacant elements in its local sub-tree.
	Then, all ranks exchange the roots of their sub-trees and construct the common upper portion of the octree.
	Whenever a rank now needs access to the children of another rank's node (e.g., when the AC is not satisfied for the branch node or it picked a remote node later on), it downloads the children and inserts them into its own tree.
	These downloaded nodes are discarded at the end of the update step so that the newly calculated values can be used in the next update step.
	Algorithm~\ref{alg:conn} shows the pseudo-code for finding targets in this parallel version.}

\begin{algorithm}
	\caption{\edit{The algorithm for finding targets for new synapses, assuming $p$ MPI ranks that work together to simulate $n$ neurons ($n/p$ per rank).
			After updating the octree, the Barnes--Hut-algorithm (lls.~10--20) is applied recursively to find partners for each neuron.
			Note here that $\Theta(\cdot)$ denotes the exact complexity class (in contrast to the $\theta$ in the acceptance criterion (AC)) and $O(\cdot)$ an upper bound.}}
	\label{alg:conn}
	\begin{algorithmic}[1]
		\State Update own leaf nodes \Comment{$\Theta(n/p)$}
		\State Update own sub-tree \Comment{$\Theta(\log (n/p))$}
		\State Exchange roots of sub-trees \Comment{$O(p)$}
		\State Update common upper portion \Comment{$\Theta(p)$}
		\For{All local neurons} \Comment{$\Theta(n/p)$ iterations}
		\State \texttt{current\_root} = \texttt{global\_root} \Comment{$\Theta(1)$}
		\While{\texttt{current\_root} is virtual} \Comment{$O(\log (n))$ iterations}
		\State \texttt{stack} $\gets$ \texttt{current\_root} \Comment{$\Theta(1)$}
		\State \texttt{nodes} $\gets \emptyset$  \Comment{$\Theta(1)$}
		\While{\texttt{stack} not empty} \Comment{$O(\log (n))$ iterations}
		\State \texttt{node} = pop first from \texttt{stack} \Comment{$\Theta(1)$}
		\For{Children \texttt{child} of \texttt{node}} \Comment{$\Theta(1)$}
		\If{\texttt{child} satisfies AC or is an actual neuron}
		\State \texttt{nodes} = \texttt{nodes} + \texttt{child} \Comment{$\Theta(1)$}
		\Else
		\State  \texttt{stack} = \texttt{stack} + \texttt{child} \Comment{$\Theta(1)$}
		\EndIf
		\EndFor
		\EndWhile
		\State \texttt{current\_root} = choose from \texttt{nodes} \Comment{$O(\log (n))$}
		\EndWhile
		\State Propose synapse to \texttt{current\_root} \Comment{$\Theta(1)$}
		\EndFor
	\end{algorithmic}
\end{algorithm}

\section*{A Sharp Runtime Bound}
\label{sec:runtime}

\begin{figure}
	\centering
	\includegraphics[width=0.8\linewidth]{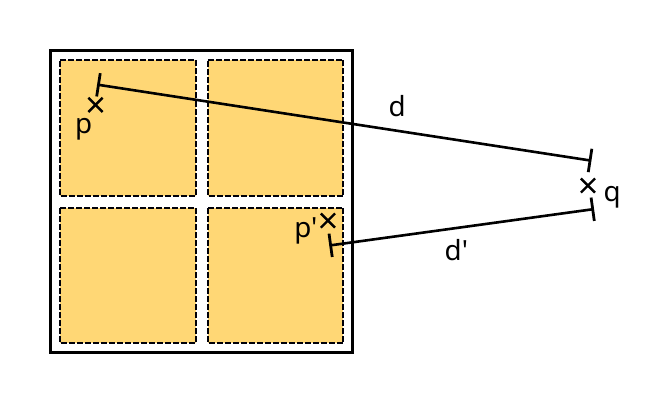}
	\caption{A neuron with position $q$ searches for a partnering neuron and the subtree with root position $p$ (white box with solid border) satisfies the Acceptance Criterion (AC). One of its children, the subtree with root position $p'$ (yellow boxes with dashed border), might not automatically satisfy (AC) because its distance $d'$ might be smaller than the distance of the parent subtree $d$.}
	\label{fig:box}
\end{figure}

In the previous publication, the authors advised setting $\theta \leq 1 / \sqrt{3}$.
This way, a subspace that contains the searching neuron is always unpacked because it never satisfies the Acceptance Criterion (AC).
This precaution prevents autapses (self connections), and a searching neuron does not have to adjust a virtual neuron's position by removing its position from the calculation.

However, the authors did not account that if a neuron selects a virtual neuron and thus needs to expand it \edit{(lls.~7--21)}, its children also probably satisfy the AC~\edit{(lls.~12--18)}, reducing the following Barnes--Hut-algorithm applications~\edit{(lls.~10--20)} to constant complexity.
We start by calculating the requirements for the children of an accepted virtual neuron to satisfy the AC. 

Assume a fixed $0 < \theta \leq 1/\sqrt{3}$, a subdomain $S$ with side lengths $l_1, l_2, l_3 \in \mathbb{R}^+$ and centroid at $p \in \mathbb{R}^3$.
Assume that the neuron which searches for a partner has position $q \in \mathbb{R}^3$, selected the subdomain $S$, and now needs to find a partner within $S$.
That is, the AC was fulfilled for $S$:
$$\frac{\max(l_1, l_2, l_3) }{ ||p - q||_2} \overset{!}{<} \theta$$
When considering any child $S'$ of $S$, its side lengths $l_1', l_2', l_3'$ are the respective halves of the side lengths of $S$.
The centroid of it can be anywhere within the original subdomain $S$---depending on which actual child is considered and where in the child's domain the centroid is.
Figure~\ref{fig:box} depicts a situation in a quadtree where one child has a smaller distance than its parent. 
For our purpose, the centroid $p'$ of the child $S'$ satisfies
$$p' = p + \epsilon $$
for some $\epsilon \in ([-l_1, l_1], [-l_2, l_2], [-l_3, l_3])$.
We can use the Cauchy--Schwarz inequality to obtain:
$$| \; ||p||_2 - ||\epsilon||_2 \; | \leq ||p'||_2 \leq ||p||_2 + ||\epsilon||_2$$
Because both $p$ and $p'$ lie within $S$, we can give a bound on $||\epsilon||_2$ as well:
$$0 \leq  ||\epsilon||_2 \leq \sqrt{l_1^2 + l_2^2 + l_3^2} \leq \sqrt{3} \cdot \max(l_1, l_2, l_3)$$
So, when asking if the child also satisfies AC, we have to establish:
$$ \frac{ 0.5 \cdot \max(l_1, l_2, l_3) }{ ||p' - q||_2} = \frac{\max(l_1', l_2', l_3') }{ ||p' - q||_2} \overset{?}{<} \theta$$
We start with the assumption and work our way to the desired result.
We set $l = \max(l_1, l_2, l_3)$ and $l' = \max(l_1', l_2', l_3')$:
\begin{align}
	l / ||p - q||_2 &\overset{!}{<} \theta \\
	l &< \theta \cdot ||p - q||_2 \\
	l < \theta \cdot ||p - q||_2 &\leq \theta \cdot (||p' - q||_2 + \sqrt{3} \cdot l) \\
	0.5 \cdot l + 0.5 \cdot l &< \theta \cdot ||p' - q||_2 + \theta \cdot \sqrt{3} \cdot l \\
	(0.5 \cdot l - \theta \cdot \sqrt{3} \cdot l) + 0.5 \cdot l &< \theta \cdot ||p' - q||_2
\end{align}
The last inequality has the general form $a + b < c$.
If we want to establish $b < c$, we have to prove $a \ge 0$, because
$$b = 0 + b \leq a + b < c$$
Thus, we now have to prove:
\begin{align}
	0.5 \cdot l - \theta \cdot \sqrt{3} \cdot l &\overset{?}{\ge} 0 \\
	0.5 \cdot l &\ge \theta \cdot \sqrt{3} \cdot l \\
	0.5 &\ge \theta \cdot \sqrt{3} \\
	1/ (2 \cdot \sqrt{3}) &\ge \theta \label{eq:theta}
\end{align}
We receive the result that whenever we have $\theta \leq 1/ (2 \cdot \sqrt{3})$, and the AC is satisfied for a virtual neuron \edit{(ll.~14)}, all of its children automatically satisfy the AC.
In this case, the Barnes--Hut-algorithm \edit{(lls.~10--20)} only needs to consider $8$ (virtual) bodies, so it has constant complexity.

We now perform the calculation again, however, we introduce a scaling factor $0 < (1/m) \leq 1$.
The calculations before do not guarantee that a child can always be approximated for $\theta > 1/ (2 \cdot \sqrt{3})$.
However, if the child itself fails the AC~\edit{(ll.~16)}, its children (the grandchildren of the subdomain that satisfied the AC) have a better chance.
We now calculate how many subdomains ($m$) the accepted subdomain would have to be split into to guarantee that all new subdomains fulfill the AC.
\begin{align}
	l < \theta \cdot ||p - q||_2 &\leq \theta \cdot (||p' - q||_2 + \sqrt{3} \cdot l) \\
	(1 - (1/m)) \cdot l + (1/m) \cdot l &< \theta \cdot ||p' - q||_2 + \theta \cdot \sqrt{3} \cdot l \\
	((1 - (1/m)) \cdot l - \theta \cdot \sqrt{3} \cdot l) + (1/m) \cdot l &< \theta \cdot ||p' - q||_2
\end{align}
Following the same reasoning as before, we get:
\begin{align}
	(1 - (1/m)) \cdot l - \theta \cdot \sqrt{3} \cdot l &\ge 0 \\
	(1 - (1/m)) \cdot l &\ge  \theta \cdot \sqrt{3} \cdot l \\
	(1 - (1/m)) &\ge \theta \cdot \sqrt{3}  \\
	(m-1)/m &\ge \theta \cdot \sqrt{3}  \\
	m - 1 &\ge \theta \cdot \sqrt{3} \cdot m \\
	m -\theta \cdot \sqrt{3} \cdot m &\ge 1 \\
	m \cdot (1 -\theta \cdot \sqrt{3}) &\ge 1 \\
	m &\ge 1 / (1 -\theta \cdot \sqrt{3}) \label{eq:m}
\end{align}
In dependence on $\theta$, the formula gives the number of subdomains the accepted virtual neuron has to be broken down into to guarantee that all of them satisfy the AC.
Because this formula does not depend on the number of approximated neurons in a subdomain, it is always a constant.
We list the values of $m$ for different values of $\theta$ in Table~\ref{tab:m} and give the upper bound on the number of nodes that need to be considered.
Therefore, after the initial choice of a partner neuron, all subsequent choices have to be made from a constant number of choices.
Thus, the overall complexity of every Barnes--Hut-approximation \edit{(lls.~10--20)} after the first one has constant runtime complexity instead of logarithmic.

As a final step in our consideration, we look at the previous bound on $||\epsilon||_2$.
In theory, the position of the centroid $p$ of a subdomain $S$ and the centroid of a child $p'$ can be anywhere in $S$.
However, if we assume that the centroid is always close to the center of its subdomain, we can improve the previous bound:
$$0 \leq  ||\epsilon||_2 \leq 0.25 \cdot \sqrt{l_1^2 + l_2^2 + l_3^2} \leq 0.25 \cdot \sqrt{3} \cdot \max(l_1, l_2, l_3)$$
Following the factor $0.25$, Equation~\ref{eq:theta} now has the form:
$$\frac{1 }{ 0.25 \cdot 2 \cdot \sqrt{3}} = \frac{2 }{ \sqrt{3} } > 1 \ge \theta$$
In this case, whenever a virtual neuron satisfies the AC, all of its children do so, too.
This does not improve the theoretical complexity; however, it sheds light on the order of over-approximation we performed during the calculations.

\begin{table}[]
	\centering
	\begin{tabular}{c|c|c|c|c}
		$\theta$    & $m$        & $\log_2(m)$ & Guaranteed AC         & Number of Nodes\\
		\hline  $0.1$       & $1.20949$  & $0.274399$   & Children              & 8 \\
		\hline  $0.2$       & $1.53001$  & $0.613541$   & Children              & 8 \\
		\hline  $0.3$       & $2.08166$  & $1.057734$   & Grandchildren         & 64 \\
		\hline  $0.4$       & $3.25542$  & $1.702844$   & Grandchildren         & 64 \\
		\hline  $0.5$       & $7.46410$   & $2.899968$   & Great-Grandchildren   & 512 
	\end{tabular}
	\caption{The evaluation of Equation~\ref{eq:m} for different values of $\theta$. Because $m$ gives the number of divisions by axis that have to happen, we calculate the 2-logarithm of $m$ to obtain an upper bound for the height in the octree.}
	\label{tab:m}
\end{table}

\section*{\edit{Complexity in the Parallel Case}}
\label{sec:parallel-case}
\edit{
	Based on the findings in the previous section, we can also update the complexity in the parallel case. 
	For each update step of the simulation, we perform Algorithm~\ref{alg:conn}; however, the number of update steps is determined before the execution, so we can ignore it for the complexity consideration and concentrate on the steps for finding the target neurons.
	Thus, in the following, we specify the complexity solely dependent on the number of neurons~$n$ and the number of MPI ranks~$p$. 
	Before the loop in line~5 begins, we have to perform updates local to each rank (lls.~1, 2, 4) with a combined complexity of $\Theta(n/p+\log(n/p)+p)$. 
	Additionally, an exchange between all ranks happens (ll.~3) with a communication complexity of $\Theta(p)$.}

\edit{The loop in line 5 is repeated for all local neurons which results in $\Theta(n/p)$ iterations. Within the loop, the Barnes--Hut-algorithm is applied repeatedly (loop at lls. 7--21), while the octree is searched for target neurons. 
	As shown in the previous section, only the first of these applications of the Barnes--Hut-algorithm (lls. 10--20) has a complexity of $\Theta(\log(n))$, all following applications only unpack a constant number of neurons, so they have constant complexity.
	Thus, for all iterations (but the first one) of the loop at lines 7--21 the number of nodes in \texttt{nodes} is constant, resulting in the selection of the current root node (ll. 20) also being constant.
	Considering this, we get a complexity of $\Theta(1 \cdot \log(n))$ (number of iterations times the complexity) for the first iteration of the target neuron search, all other iterations together have a complexity of $\Theta(\log(n) \cdot 1)$ (again, number of iterations times the complexity). 
	Taking the repetition for each neuron (loop at ll. 5) into account, we get a complexity of $\Theta((1 \cdot \log(n) + \log(n) \cdot 1)\cdot n/p) = \Theta(n/p \cdot \log(n))$ and including lines 1--4, we get $\Theta(n/p \cdot \log(n) + p)$.} 

\edit{In the previous publication, the authors tested the algorithm and fitted performance models for their weak-scaling experiments with different acceptance criteria (for $\theta = 0.3$, they calculated $0.961461 + 0.14743 \cdot \log_2(p)$; for $\theta = 0.4$, they calculated $0.415784 + 0.0652235 \cdot \log_2(p)$).
	This allows us---for practical application---to drop ``$+p$'' from the complexity class.
	Lines 3 and 4 of Algorithm~\ref{alg:conn} have no practical runtime implications, the associated work is just too small, even in the case of $2^{18}$ MPI ranks.
	Using this insight and setting $n = p \cdot 5000$ (5000 neurons per MPI rank as in their experiments) in the complexity class, we see that their findings support our claim: $\Theta(n/p \cdot \log(n)) = \Theta((p  \cdot 5000)/p \cdot \log(p  \cdot 5000)) = \Theta(\log(p))$.}

\section*{Conclusion}
In this comment paper, we have shown that the previous runtime \edit{complexity} analysis of the adapted Barnes--Hut-algorithm was too conservative.
When searching for a partner neuron out of $n$ possible, the first (full) Barnes--Hut-approximation has a runtime complexity of $O(\log(n))$.
However, the following $O(\log(n))$ calculations have constant runtime complexity.
Overall, finding suitable partners for all neurons has runtime complexity $O(n \cdot (\log(n) + \log(n))) = O(n \cdot \log(n))$.
This analysis carries easily over to the parallel case, reducing the runtime complexity to $O(n/p \cdot \log(n))$.

In the previous publication, the authors tested the algorithm with up to $1\,310\,720\,000$ neurons.
If we assume a balanced octree, its height would be $10 \approx \log_8(1\,310\,720\,000)$, which in practice is not too far away from $8$, showing that this improvement is only of theoretical nature.

\section*{Acknowledgments} 
We acknowledge the support of the European Commission and the German Federal Ministry of Education and Research (BMBF) under the EuroHPC Programme DEEP-SEA (955606, BMBF Fund-ing No. 16HPC015). The EuroHPC Joint Undertaking (JU) receives support from the European Union’s Horizon 2020 research and in-novation programme and GER, FRA, ESP, GRC, BEL, SWE, UK, CHE. This research was also supported by the EBRAINS research infras-tructure, funded by the European Union’s Horizon 2020 Framework Programme for Research and Innovation under the Specific GA No. 945539 (Human Brain Project SGA3), and is partly funded by the Federal Ministry of Education and Research (BMBF) with Specific GA No. NHR2021HE and the State of Hesse as part of the NHR Pro-gram under ``Kapitel 1502, Förderprodukt 19 NHR4CES''.

\bibliographystyle{plain}
\bibliography{references}
\end{document}